\documentclass[twocolumn]{aastex63}

\received{September 29, 2020}
\revised{January 8, 2021}
\accepted{January 22, 2021}
\submitjournal{ApJL}

\shorttitle{Shadows, spirals and dust tails in SU~Aur}
\shortauthors{Ginski et al.}
\graphicspath{{./}{figures/}}

\begin{document}
\title{Disk Evolution Study Through Imaging of Nearby Young
Stars (DESTINYS): \\ Late infall causing disk misalignment and dynamic structures in SU Aur\footnote{Based on observations performed with VLT/SPHERE under program ID 1104.C-0415(E)}}

\correspondingauthor{Christian Ginski}
\email{c.ginski@uva.nl}

\author[0000-0002-4438-1971]{Christian Ginski}
\affiliation{Anton Pannekoek Institute for Astronomy, University of Amsterdam, Science Park 904, 1098XH Amsterdam, The Netherlands}
\affiliation{Leiden Observatory, Leiden University, PO Box 9513, 2300 RA Leiden, The Netherlands}

\author[0000-0003-4689-2684]{Stefano Facchini}
\affiliation{European Southern Observatory, Karl-Schwarzschild-Strasse 2, 85748 Garching bei München, Germany}

\author[0000-0001-6947-6072]{Jane Huang}
\affiliation{Department of Astronomy, University of Michigan, 323 West Hall, 1085 S. University Avenue, Ann Arbor, MI 48109, USA}
\affiliation{NHFP Sagan Fellow}

\author[0000-0002-7695-7605]{Myriam Benisty}
\affiliation{Unidad Mixta Internacional Franco-Chilena de Astronom\'ia, CNRS, UMI 3386 and Departamento de Astronom\'ia, Universidad de Chile, Camino El Observatorio 1515, Las Condes, Santiago, Chile}
\affiliation{Univ. Grenoble Alpes, CNRS, IPAG, 38000 Grenoble, France}

\author{Dennis Vaendel}
\affiliation{Leiden Observatory, Leiden University, PO Box 9513, 2300 RA Leiden, The Netherlands}

\author[0000-0001-9524-3408]{Lucas Stapper}
\affiliation{Anton Pannekoek Institute for Astronomy, University of Amsterdam, Science Park 904, 1098XH Amsterdam, The Netherlands}

\author[0000-0002-3393-2459]{Carsten Dominik}
\affiliation{Anton Pannekoek Institute for Astronomy, University of Amsterdam, Science Park 904, 1098XH Amsterdam, The Netherlands}

\author[0000-0001-7258-770X]{Jaehan Bae}
\affiliation{Department of Terrestrial Magnetism, Carnegie Institution for Science, 5241 Broad Branch Road NW, Washington, DC 20015, USA}
\affiliation{NHFP Sagan Fellow}

\author[0000-0002-1637-7393]{Fran\c{c}ois M\'enard}
\affiliation{Univ. Grenoble Alpes, CNRS, IPAG, 38000 Grenoble, France}

\author[0000-0002-8670-2809]{Gabriela Muro-Arena}
\affiliation{Anton Pannekoek Institute for Astronomy, University of Amsterdam, Science Park 904, 1098XH Amsterdam, The Netherlands}

\author[0000-0001-5217-537X]{Michiel R. Hogerheijde}
\affiliation{Leiden Observatory, Leiden University, PO Box 9513, 2300 RA Leiden, The Netherlands}
\affiliation{Anton Pannekoek Institute for Astronomy, University of Amsterdam, Science Park 904, 1098XH Amsterdam, The Netherlands}

\author[0000-0003-1878-327X]{Melissa McClure}
\affiliation{Leiden Observatory, Leiden University, PO Box 9513, 2300 RA Leiden, The Netherlands}

\author[0000-0003-1520-8405]{Rob G. van Holstein}
\affiliation{European Southern Observatory (ESO), Alonso de C\'ordova 3107, Vitacura, Casilla 19001, Santiago de Chile, Chile}
\affiliation{Leiden Observatory, Leiden University, PO Box 9513, 2300 RA Leiden, The Netherlands}

\author[0000-0002-1899-8783]{Tilman Birnstiel}
\affiliation{University Observatory, Faculty of Physics, Ludwig-Maximilians-Universität München, Scheinerstr. 1, D-81679 Munich, Germany}
\affiliation{Exzellenzcluster ORIGINS, Boltzmannstr. 2, D-85748 Garching, Germany}

\author[0000-0002-8692-8744]{Yann Boehler}
\affiliation{Rice University, Department of Physics and Astronomy, Main Street, 77005 Houston, USA}

\author[0000-0003-1401-9952]{Alexander Bohn}
\affiliation{Leiden Observatory, Leiden University, PO Box 9513, 2300 RA Leiden, The Netherlands}

\author[0000-0002-9298-3029]{Mario Flock}
\affiliation{Max Planck Institute for Astronomy, Königstuhl 17, 69117 Heidelberg, Germany}

\author[0000-0003-2008-1488]{Eric E. Mamajek}
\affiliation{Jet Propulsion Laboratory, California Institute of Technology, M/S 321-100, 4800 Oak Grove Drive, Pasadena, CA 91109, USA}

\author[0000-0003-3562-262X]{Carlo~F. Manara}
\affiliation{European Southern Observatory, Karl-Schwarzschild-Strasse 2, 85748 Garching bei München, Germany}

\author[0000-0001-8764-1780]{Paola Pinilla}
\affiliation{Max Planck Institute for Astronomy, Königstuhl 17, 69117 Heidelberg, Germany}

\author[0000-0001-5907-5179]{Christophe Pinte}
\affiliation{School of Physics and Astronomy, Monash University, Clayton, Vic 3800, Australia}

\author[0000-0003-3133-3580]{\'{A}lvaro Ribas}
\affiliation{European Southern Observatory (ESO), Alonso de C\'ordova 3107, Vitacura, Casilla 19001, Santiago de Chile, Chile}



\begin{abstract}
Gas-rich circumstellar disks are the cradles of planet formation. As such, their evolution will strongly influence the resulting planet population.
In the ESO DESTINYS large program, we study these disks within the first 10\,Myr of their development with near-infrared scattered light imaging. \\
Here we present VLT/SPHERE polarimetric observations of the nearby class II system SU\,Aur in which we resolve the disk down to scales of $\sim$ 7\,au.
In addition to the new SPHERE observations, we utilize VLT/NACO, HST/STIS and ALMA archival data.\\
The new SPHERE data show the disk around SU\,Aur and extended dust structures in unprecedented detail. We resolve several dust tails connected to the Keplerian disk. By comparison with ALMA data, we show that these dust tails represent material falling onto the disk. 
The disk itself shows an intricate spiral structure and a shadow lane, cast by an inner, misaligned disk component.\\
Our observations suggest that SU\,Aur is undergoing late infall of material, which can explain the observed disk structures. SU\,Aur is the clearest observational example of this mechanism at work and demonstrates that late accretion events can still occur in the class II phase, thereby significantly affecting the evolution of circumstellar disks. Constraining the frequency of such events with additional observations will help determine whether this process is responsible for the spin-orbit misalignment in evolved exoplanet systems.  

\end{abstract}



\section{Introduction}

Circumstellar disks are the birth places of planetary systems. Thus their physical properties strongly influence the outcome of the planet formation processes. In turn, massive planets dramatically impact the disk structure. Recent scattered light observations have revealed a number of disks with warps or misalignments of inner and outer disk regions (e.g., \citealt{Marino2015, Benisty2017}). The interaction with either planets or stellar companions is frequently invoked to explain these observations (e.g., \citealt{2012Natur.491..418B, Facchini2018}). Recently \cite{Bi2020} and \cite{Kraus2020} showed multiple misaligned rings supporting this scenario in the GW\,Ori triple system.
However, many of the systems with inferred misalignments are around single stars (e.g., \citealt{Pinilla2018, Muro-Arena2020}). 
Similarly, spin-orbit misalignment of transiting planets possibly inherited from the gas-rich disk phase, is common in single stellar systems
(e.g., \citealt{Triaud2010}). An alternative scenario is that disk misalignments are natural consequence of angular momentum transfer due to late infall of material on the disk \citep{Thies2011,Dullemond2019}. Observations probing both large and small spatial scales have the potential to test this possibility. \\  
SU\,Aur is a nearby (158.4$\pm$1.5\,pc, \citealt{GAIA-DR2-2018}) classical T Tauri star in the Taurus-Auriga star forming region. Spectroscopic studies determined a spectral type of G4 and a stellar luminosity of $log(L/L_\odot) = 0.9$ \citep[when re-scaled with the latest distance estimate;][]{Herczeg2014}.
Using the spectral type as well as this luminosity as input for stellar model isochrones \citep{Siess2000}, we find a stellar mass of 2.0$^{+0.2}_{-0.1}\,M_\odot$ and an age range of 4-5.5\,Myr.\\ 
SU\,Aur is surrounded by extended circumstellar structure first resolved in near infrared scattered light \citep{Jeffers2014}.  
The signal extends up to 500\,au and shows a strong asymmetry along the East-West direction. Subsequent scattered light observations detected a faint dust tail extending from the main disk toward the West \citep{deLeon2015}. 
A strong azimuthal brightness asymmetry is attributed to the dust scattering phase function and to a higher surface density on the northern side of the disk. ALMA observations show a Keplerian disk and a gas tail that extends out to several hundreds of au to the West \citep{Akiyama2019}, that could either be caused by a disruption of the disk, by a perturber, or trace cloud material accreting onto the disk.\\
In this letter, we present new observations of SU\,Aur obtained as part of the DESTINYS program (Disk Evolution Study Through Imaging of Nearby Young Stars \citealt{Ginski2020}) that aims to study the circumstellar environment of nearby T Tauri stars, complemented with VLT/NACO, HST/STIS and ALMA archival data. %


\section{Observations}

We obtained new observations of SU\,Aur with VLT/SPHERE (\citealt{Beuzit2019}), and use archival data taken with VLT/NACO  (program ID: 088.C-0924, PI: S. Jeffers) and ALMA (program ID: 2013.1.00426.S, PI: Y. Boehler).

\subsection{SPHERE observations}

SU\,Aur was observed on 14th of December 2019 with SPHERE/IRDIS in dual-beam polarimetric imaging mode (DPI, \citealt{deBoer2020,vanHolstein2020})  in the broad band H filter with pupil tracking setting. The central star was placed behind an apodized Lyot coronagraph with an inner working angle\footnote{IWA defined as the separation at which the throughput reaches 50\%.} of 92.5\,mas. The individual frame exposure time was set to 32\,s and a total of 104 frames where taken separated in 26 polarimetric cycles of the half wave plate. The total integration time was 55.5\,min. Observations conditions were excellent with an average Seeing of 0.8\arcsec and an atmosphere coherence time of 6.7\,ms.\\
The data were reduced using the public IRDAP pipeline (IRDIS Data reduction for Accurate Polarimetry, \citealt{vanHolstein2020}). The images were astrometrically calibrated using the pixel scale and true north offset given in \cite{2016SPIE.9908E..34M}. \\
Since the data were taken in pupil tracking mode we were able to perform angular differential imaging (ADI, \citealt{2006ApJ...641..556M}) reduction in addition to the polarimetric reduction, resulting in a total intensity image and a polarized intensity image. We show the final result of both post-processing approaches in figure~\ref{fig:sphere_images}.
Note that instead of polarized intensity we show the radial Stokes parameter Q$\phi$ as is now standard in most studies. We follow the definition in \cite{deBoer2020}:

\begin{equation}
    Q_\phi = -Q\,cos(2\phi) - U\,sin(2\phi)
\end{equation}

\subsection{NACO observations}

\begin{figure*}
\center
\includegraphics[width=0.68\textwidth]{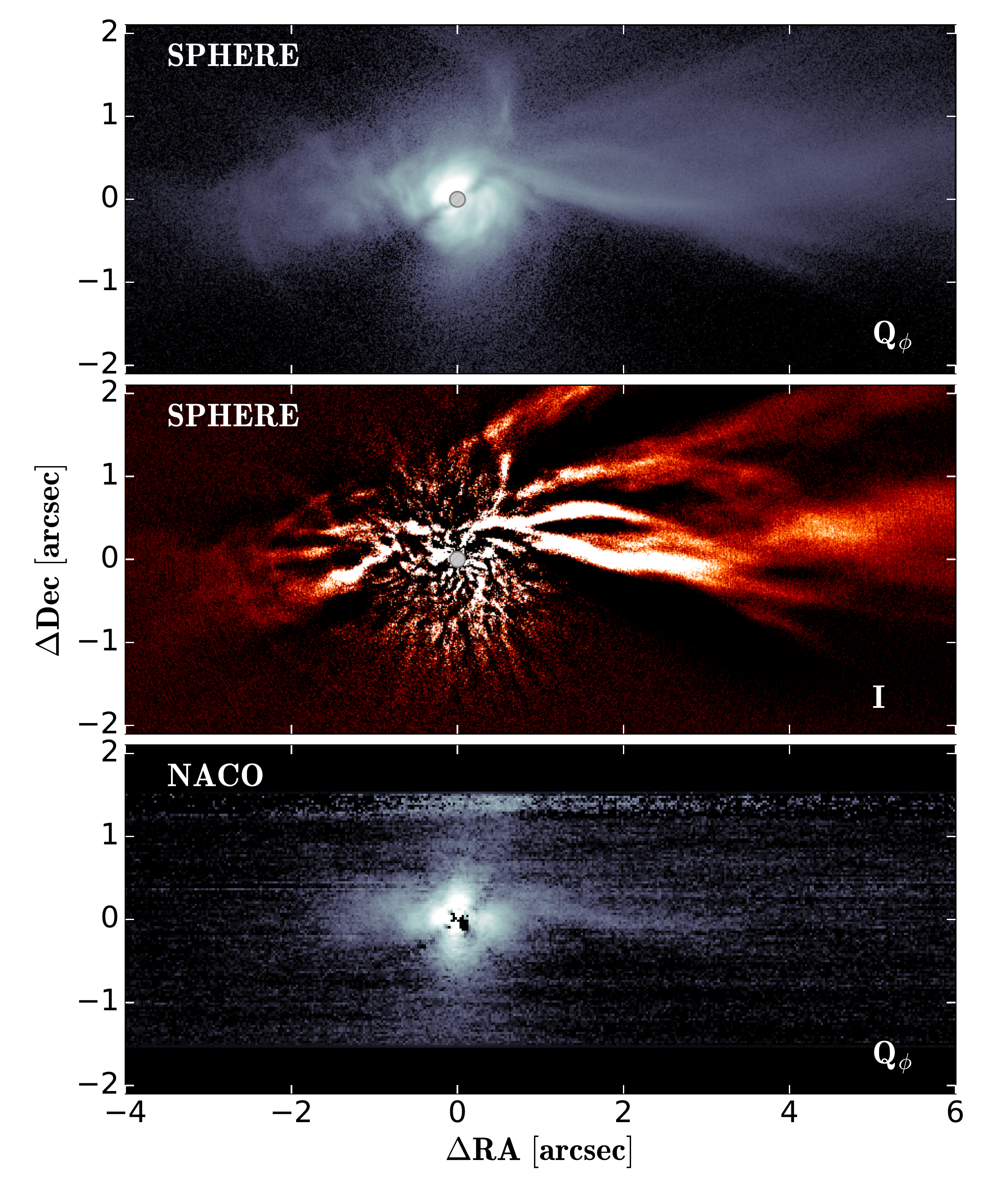} 
\caption{SPHERE/IRDIS and NACO observations of the SU\,Aur system. We show SPHERE H-band Q$_\phi$ polarized signal on the top, SPHERE H-band total intensity in the center panel and NACO K$_S$-band Q$_\phi$ polarized signal on the bottom. 
The SPHERE coronagraph is marked with a hashed circle.
} 
\label{fig:sphere_images}
\end{figure*}

SU\,Aur was observed with VLT/NACO in polarimetric imaging mode on 2nd of November 2011 in the Ks filter. Observing conditions were fair with an average Seeing of 0.8\arcsec{} and a coherence time of 3\,ms. As NACO does not offer a coronagraph in polarimetric mode, short individual frame exposure times of 0.35\,s were used. A total of 8160 frames were taken separated in 24 polarimetric cycles. This amounts to a total integration time of 47.6\,min. The data was taken in dithering mode in order to allow for sky background subtraction. \\
The data reduction was performed in principle analogous to the SPHERE data, however without the benefit of a detailed instrument model to determine instrument polarization. The instrumental polarization was thus estimated from the data, by placing a small aperture at the central star location and with an aperture radius smaller than one resolution element, i.e. where we would expect the polarimetric signal to be unresolved and thus on average small. Other data reduction steps were performed as described in \cite{Ginski2016} for an analogous data set of HD\,97048. The resulting Q$\phi$ image is shown in figure~\ref{fig:sphere_images}.

\subsection{Archival ALMA observations}
We retrieved CO and continuum data of SU Aur observed as part of program 2013.1.00426.S (PI: Y. Boehler) from the ALMA archive.
The 880 $\mu$m continuum and $^{12}$CO $J=3-2$ line were observed in Band 7 on 2015 July 24 and 2016 July 23 with 44 and 43 antennas, respectively. For the first set of observations, baselines ranged from 15 to 1600 m and the quasars J0423-0120, J0423-013, and J0438+3004 served as the bandpass, amplitude, and phase calibrators, respectively. For the second set of observations, the baselines ranged from 17 to 1100 m and the quasars J0510+1800, J0238+1636, and J0433+2905 served as the bandpass, amplitude, and phase calibrators, respectively. The correlator was set up with four  spectral windows (SPWs). The SPW covering the CO $J=3-2$ line was centered at 345.797 GHz and had a bandwidth of 468.750 MHz and channel widths of 122.07 kHz. The other three SPWs were centered at 334.016, 335.974, and 347.708 GHz, and each one had a bandwidth of 2 GHz and channel widths of 15.625 MHz.  The cumulative on-source integration time was 9 minutes.  

The 1.3 mm continuum, $^{12}$CO $J=2-1$, and $^{13}$CO $J=2-1$ were observed in Band 6 on 19 July 2015 and 8 August 2015. Observational details are provided in \citet{Akiyama2019}, which first published the 1.3 mm continuum and $^{12}$CO $J=2-1$ data. In this work, we restrict our focus to the $^{13}$CO $J=2-1$ line from the Band 6 dataset, since the S/N and spatial resolution of the Band 7 continuum and $^{12}$CO line are better than their Band 6 counterparts.

The SU Aur data downloaded from the archive were processed with the ALMA pipeline in CASA v. 4.5.3. Subsequent self-calibration and imaging took place in CASA v. 5.4.0. Channels with line emission were flagged and the SPWs were spectrally averaged to form continuum datasets. The $\texttt{fixvis}$ and $\texttt{fixplanets}$ tasks were using to align the continuum peak positions of the execution blocks within each band and to assign a common phase center, respectively. One round of phase-self calibration was applied to the continuum data for the separate bands. The self-calibration solutions were then applied to the full-resolution data. The \texttt{uvcontsub} task was used to subtract the continuum from the line spectral windows in the $uv$ plane. 

A Briggs robust value of 0.5 was used with the Clark CLEAN algorithm to produce the final 880 $\mu$m continuum image, which has a synthesized beam of $0.29''\times0.17''$ (26.4$^\circ$) and an rms of 0.1 mJy beam$^{-1}$. A Briggs robust value of 1.0 was used with the multiscale CLEAN algorithm to produce the $^{12}$CO $J=3-2$ image cube, which was then primary beam-corrected with \texttt{impbcor}. The resulting syntheized beam is $0.32''\times0.19''$ (26.4$^\circ$) and the rms is 10 mJy beam$^{-1}$ in channels 0.25 km s$^{-1}$ wide. A Gaussian $uv$ taper and a Briggs robust value of 2.0 were applied to the weaker $^{13}$CO $J=2-1$ line in order to improve sensitivity. The synthesized beam is  $0.53''\times0.46''$ (15.0$^\circ$) and the rms is 10 mJy beam$^{-1}$ in channels 0.25 km s$^{-1}$ wide.

\section{Morphology in scattered light}
As evidenced from figure~\ref{fig:sphere_images}, SU Aur shows a complex circumstellar environment, with a disk, a dark lane, and large scale features. These features are discussed in details in the following.

\subsection{The circumstellar disk}
\label{sec: scatterd-light disk}
\begin{figure*}
\center
\gridline{\fig{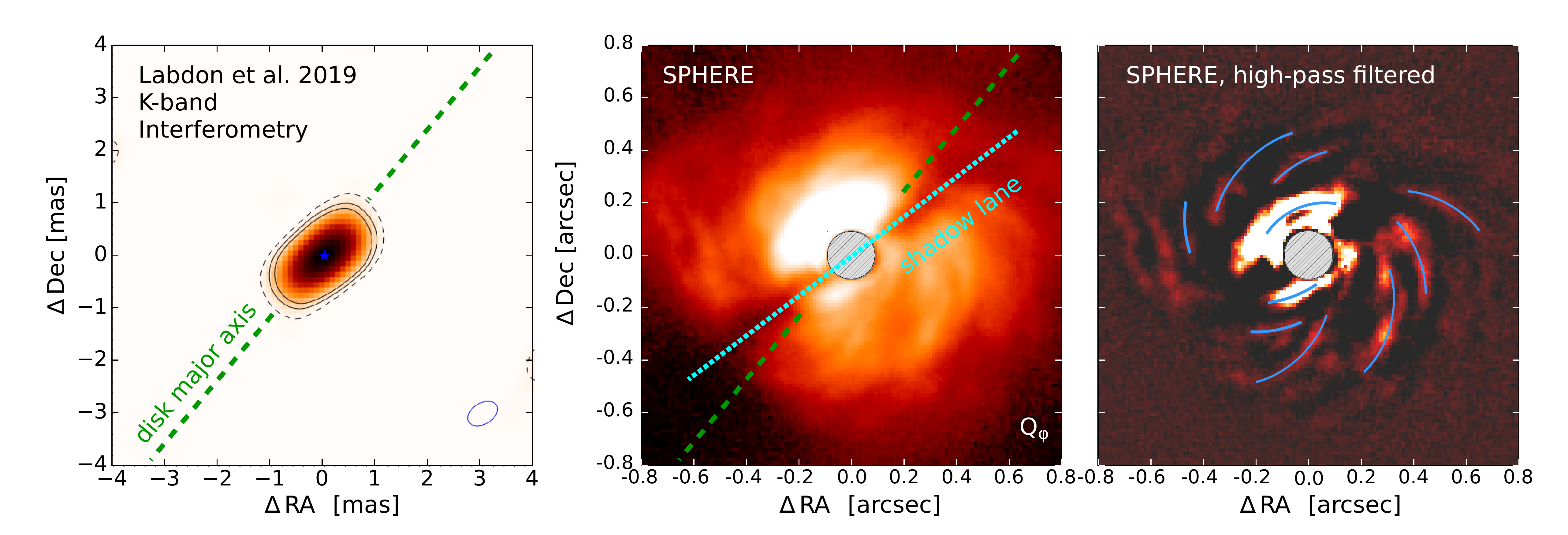}{0.99\textwidth}{}}
\gridline{\fig{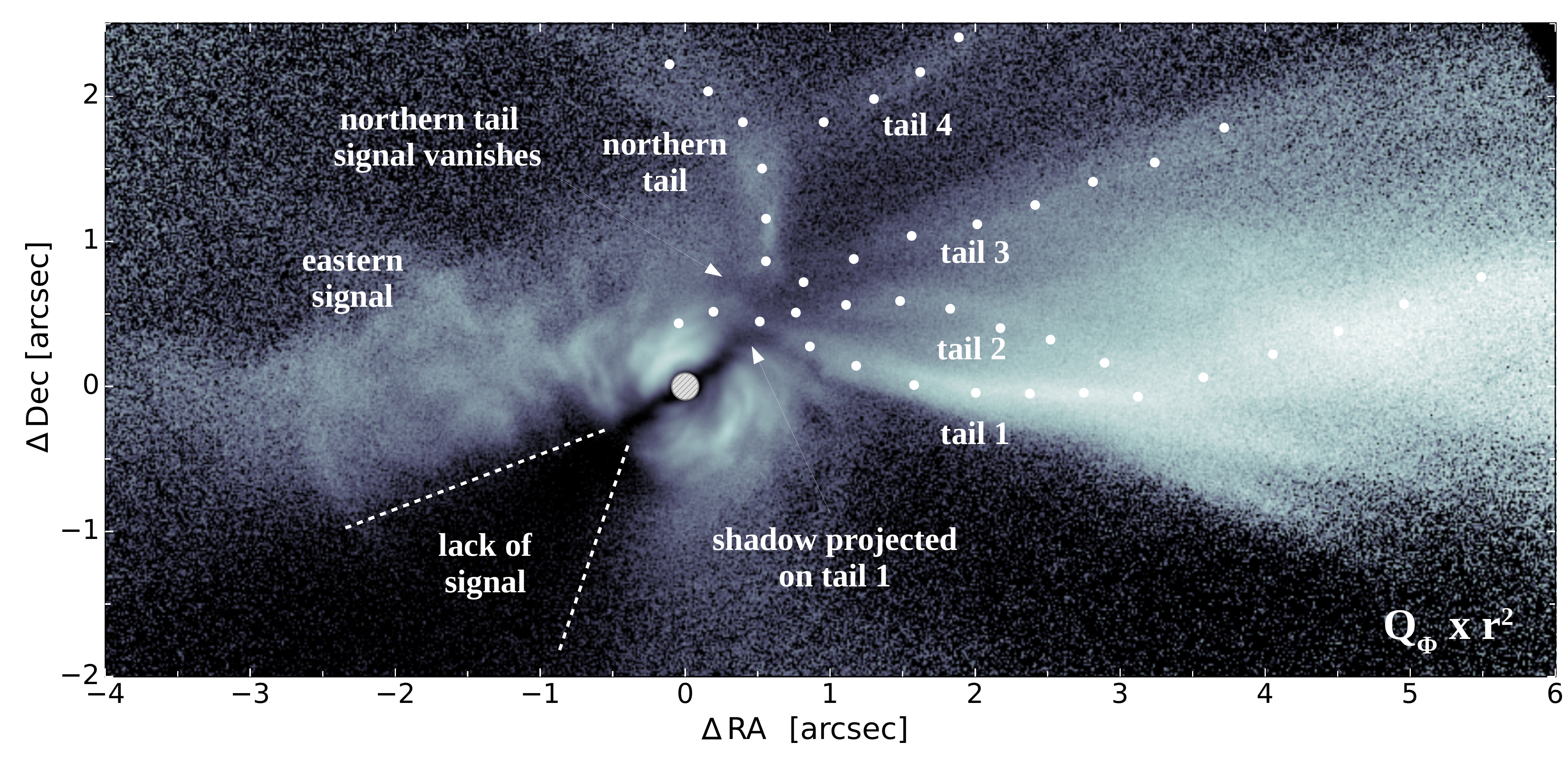}{0.98\textwidth}{}} 
\caption{
\emph{Upper-Left:} Near-infrared interferometric image reconstructed by \cite{Labdon2019} (reproduction of their figure~3). We overlay the major axis of the inner disk that they recover. \emph{Upper-Middle:} SPHERE/IRDIS Q$_\phi$ image of the innermost part of SU\,Aur. We overlay the major axis of the interferometric inner disk and the direction of the shadow lane seen in scattered light on the outer disk. \emph{Upper-Right:} Same as the middle panel, but after application of a high pass filter. We mark the visible spiral structures.
\emph{Bottom:} SPHERE Q$_\phi$ image of SU\,Aur. Several large scale structures are annotated. The polarized flux was scaled by the inclination corrected square of the separation from the primary star to compensate for the illumination drop-off.
} 
\label{fig:sphere-tails}
\end{figure*}

The most striking morphological feature is a dark lane at a position angle of $\sim$125$^\circ$, as indicated in the upper-center panel of figure~\ref{fig:sphere-tails}. This dark lane is also detected in Subaru/HiCIAO polarimetric observations \citep{deLeon2015} and in the NACO Ks-band (figure~\ref{fig:sphere_images}, bottom panel), and could trace shadowing by a misaligned inner disk.
While the extent of the disk is not directly evident from our SPHERE observations, the ALMA observations indicate the presence of a circumstellar disk in the continuum, and in the velocity channels of the $^{12}$CO $J=3-2$ line that show Keplerian rotation. The scattered light signal that is co-spatial with this kinematic signature is detected out to $\sim$0.7\arcsec\,(111\,au). The inner disk (within 1\,au) was resolved with near infrared interferometry by \cite{Labdon2019}. With image reconstruction they found it to be inclined by 52.8$^\circ\pm$2$^\circ$ with a position angle of 140.1$^\circ\pm$0.2$^\circ$, and the near side towards the West\footnote{We note that we use the convention that at a position angle of 0$^\circ$ the near side of a hypothetical disk is located to the West. In this convention the position angle reported by \cite{Labdon2019} is 320.1$^\circ$.}.
We show the reconstructed image from the interferometric data in figure~\ref{fig:sphere-tails}, upper-left panel. The inner disk position angle is close to the position angle of the dark lane seen in the outer disk.
We compared the position angle of the dark lane between the archival NACO data taken in 2011, the literature Subaru data, taken in 2014, and the new SPHERE data taken in 2019. Within this $\sim$8\,year timeframe we do not find a significant\footnote{The calibration accuracy of SPHERE is on the order of 0.1$^\circ$, while for NACO it is on the order of 0.2$^\circ$.} change in the orientation. A change in orientation would be expected if the inner disk misalignment was caused by mutual interactions with a short period binary companion, a scenario found to be unlikely.\\
We find that the disk is significantly brighter in the North-East than in the South-West. This was already reported by \cite{deLeon2015}, who interpreted this as an azimuthally asymmetric dust distribution. However this brightness asymmetry would be a natural consequence of the scattering phase function if we see an inclined disk with the near side in the North-East. Indeed from the ALMA data discussed in section~\ref{section: ALMA comparion}, we find a most likely position angle of 122.9$^\circ\pm$1.2$^\circ$ for the outer disk, which fits well with this interpretation. Given this position angle and assuming that the North-East side is the near side of the outer disk, this implies a strong misalignment between the inner and outer disk. Using the values provided by \cite{Labdon2019} for the inner disk and our measurement for the outer disk we find a relative inclination of $\sim$70$^\circ$. This is consistent with the presence of a narrow shadow lane. \\
In addition to the brightness asymmetry, we see considerable sub-structure in the disk. In particular several spiral features are present. In figure~\ref{fig:sphere-tails}, upper-right panel, we show the high-pass filtered version of the SPHERE Q$_\phi$ image with the spiral features highlighted. We can clearly identify six features in the South-West and four features in the North-East, with possible other (not highlighted) features in the East. The spirals in the South-West have generally larger pitch angles (25$^\circ$ to 49$^\circ$) than the spirals in the North-East (7$^\circ$ to 22$^\circ$). This could be explained by projection effects on an inclined scattering surface if  the North-East side is indeed the front side of the disk \citep{Dong2016}. 

\subsection{The extended structure}
\label{scattering-tail-section}

\begin{figure}
\center
\includegraphics[width=0.48\textwidth]{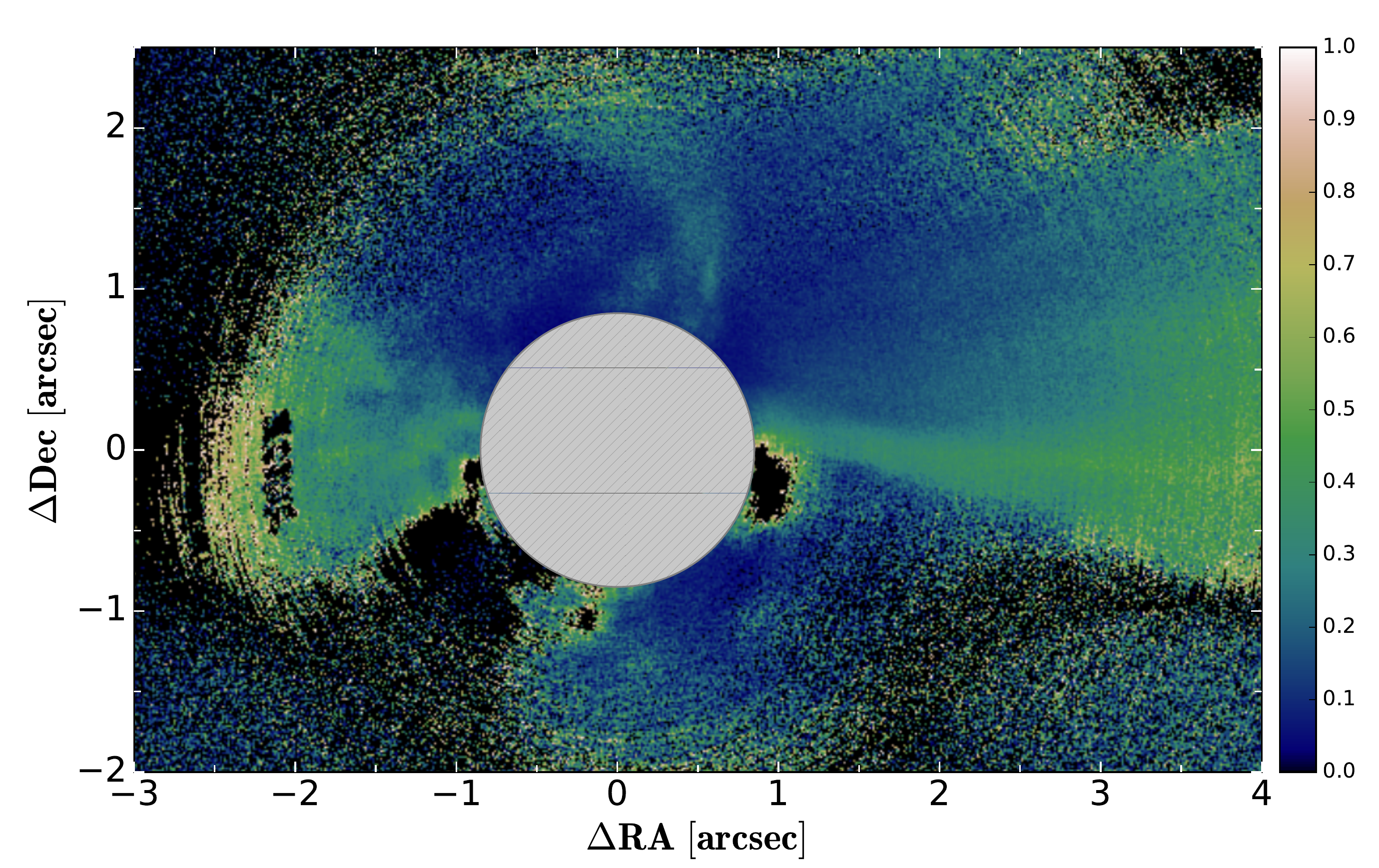} 
\caption{SPHERE degree of linear polarization. The total intensity image used is the iterative RDI reduction shown in appendix~\ref{IRDI-IADI}. Regions for which the total intensity values were not well recovered were set to 0. The inner disk is masked since it is dominated by artefacts in total intensity.
} 
\label{fig:degree-of-pol}

\end{figure}

\cite{deLeon2015} reported the detection of a dust tail with Subaru/HiCIAO in scattered polarized light extending from the disk around SU\,Aur roughly 2.5\arcsec{} to the West. We find a similar tail structure in both SPHERE and NACO data. The lower signal-to-noise NACO data show a single tail, while the SPHERE data show that this structure extends much further out than seen with either HiCIAO or NACO and consists of several tails with different curvatures and orientations. 
In figure~\ref{fig:sphere-tails} the most prominent structures are annotated. There are at least 4 distinct tails that extend towards the West labeled 1 to 4. The brightest tail in the SPHERE Q$_\phi$ image is the southern-most of these structures, i.e., tail 1. This tail is the structure seen in the NACO data and in the Subaru data. In the SPHERE image it becomes clear that this tail connects with the northern part of the Keplerian disk. Not only can we trace the tail structure until it merges with the disk, but we can also see the projection of the shadow lane from the outer disk on the dust tail. (see annotation in figure~\ref{fig:sphere-tails}). The angle of the shadow changes as would be expected if the dust tail is approaching the disk from above the disk-plane, i.e., from in between the observer and the disk.\\  
We see several fainter tails located north of tail 1, marked with numbers 2-4. Some of them are more visible in the total intensity ADI image shown in figure~\ref{fig:sphere_images}, middle panel, indicating that they likely have a low degree of polarization. We also see a structure extending to the north at a significantly different angle than tail 1-4, and labeled northern tail in figure~\ref{fig:sphere-tails}. 
The northern tail appears to vanish just before it reaches the disk (see annotation in figure~and the n\ref{fig:sphere-tails}), indicating that it is either below tail 1 or behind the disk. In order to better understand the geometry of the system we computed the degree of linear polarization of the extended structures. We utilized an iterative reference differential imaging approach (Vaendel et al. in prep.), complemented with angular differential imaging (Stapper et al., in prep.), briefly described in appendix~\ref{IRDI-IADI}.
The result is shown in figure~\ref{fig:degree-of-pol}.
Both tail 1 and the northern tail stick out with a higher degree of linear polarization compared to the surrounding structures. Assuming a standard bell curve to map the degree of linear polarization to the scattering angles (e.g., \citealt{Stolker2016b}), both tail 1 and the northern tail should be at intermediate scattering angles, with an ambiguity between forward and back-scattering. However, tail 1 is significantly brighter in the SPHERE Q$_\phi$ image than the northern tail (factor 1.5 to 4 depending on the point of measurement). This is also evidenced by the fact that tail 1 is detected by SPHERE, NACO and Subaru, whereas the northern tail is only visible in the highest signal-to-noise SPHERE data. Given that tail 1 can be smoothly traced until it connects with the northern part of the disk (i.e., the near side), it is clear that the light from tail 1 is scattered with angles smaller than 90$^\circ$. Since the northern tail shows a similar degree of polarization, but overall smaller signal, we conclude that light is scattered with angles larger than 90$^\circ$.
This means the northern tail should be located behind the disk along the line of sight. \\ 
In addition to the distinct tail-like structures, we see a more complex signal to the East of the disk. Followed by a zone where we see a distinct lack of signal toward the South-East (see annotation in figure~\ref{fig:sphere-tails}). If the signal to the East and South-East is located above the Keplerian disk (i.e., closer to the observer), then the region without signal might be a natural continuation of the shadow lane visible on the disk. In particular the signal to the East shows a very similar degree of linear polarization to tail 1, indicating similar scattering angles.

\section{ALMA observations}
\label{section: ALMA comparion}

\begin{figure}
\center
\includegraphics[width=0.48\textwidth]{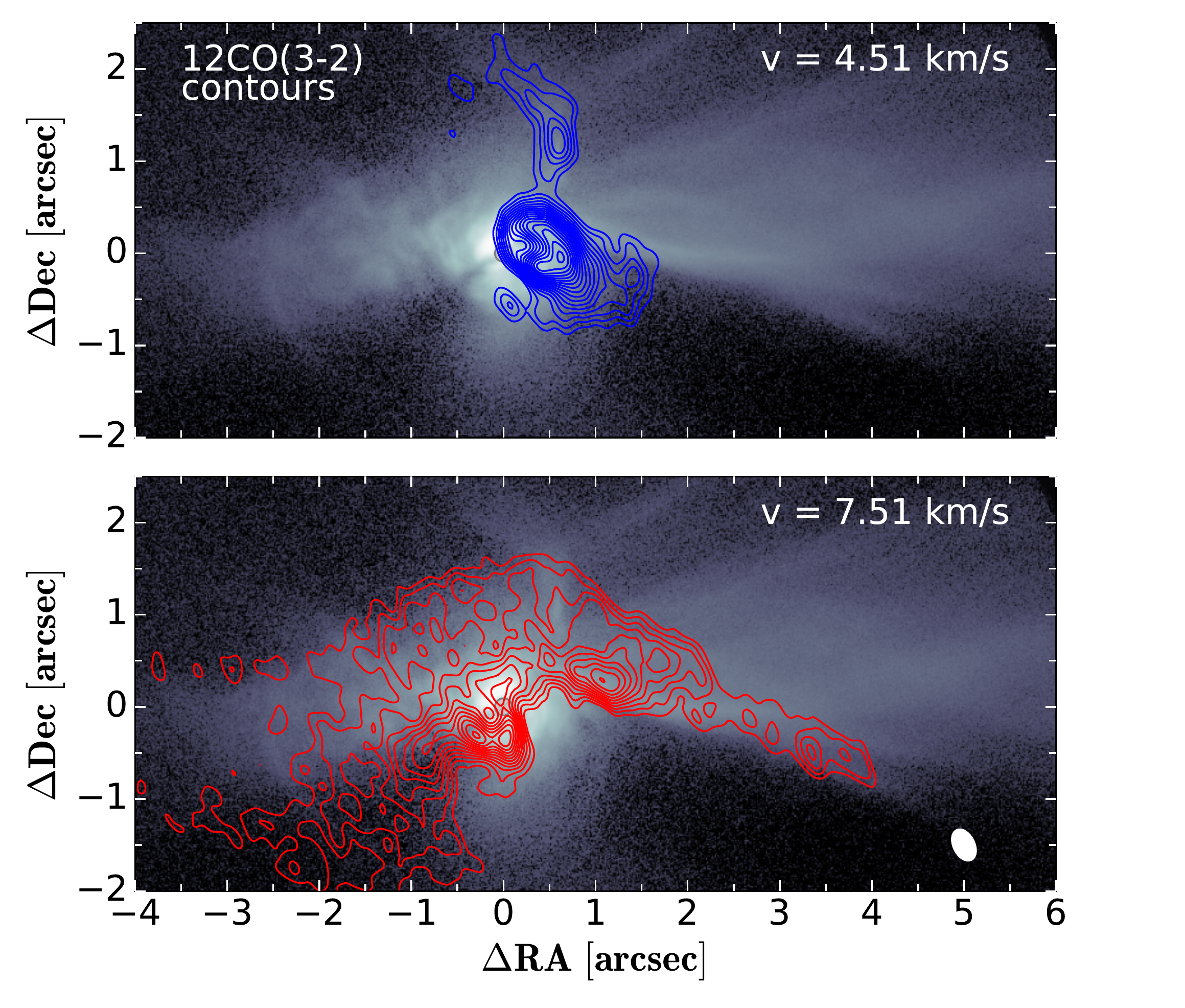} 
\caption{SPHERE Q$_\phi$ image with the ALMA $^{12}$CO 3-2 channel maps as blue and red contour overlay. Contour levels correspond to the 2\,$\sigma$ rms levels starting at 5\,$\sigma$. The ALMA beam size and orientation is indicated by the white ellipse in the lower right corner. The upper panel corresponds to signal blue-shifted (-1.49kms$^{-1}$) relative to the systemic velocity and the bottom panel corresponds to red-shifted signal (+1.51km$s^{-1}$). 
} 
\label{fig:sphere-alma-overlay}
\end{figure}

\cite{Akiyama2019} presented ALMA Band 6 observations of SU Aur, showing the dust continuum emission of the Keplerian disk and revealing an extended tail structure to the West in the gas.
The dust continuum emission shows a marginally resolved disk without particular features. In figure~\ref{fig:sphere-alma-dust} we show an overlay of the mm continuum emission and the SPHERE data. Large dust particles are concentrated at the location of the disk also seen with SPHERE but are not detected in the tail structures to the West. The mm-emission appears less extended than the scattered light, possibly indicating efficient radial drift of the large dust particles.\\
From a fit of a simple symmetric model to the continuum emission, we find a position angle of 122.9$^\circ\pm$1.2$^\circ$ and an inclination of 53.0$^\circ\pm$1.5$^\circ$ (see appendix~\ref{app:alma-fit} for details). Given that the gas and scattered light show a highly asymmetric structure it is possible that this fit is affected by systematic uncertainties.\\
In figure~\ref{fig:sphere-alma-overlay} we show an overlay of two velocity channels of the $^{12}$CO 3-2 line emission with the SPHERE data.  The first channel corresponds to a velocity of $\sim$4.51\,km\,s$^{-1}$ and is blue shifted relative to the intrinsic system velocity \citep[$\sim$6\,km\,s$^{-1}$;][]{Akiyama2019}, while the second channel corresponds to a velocity of $\sim$7.51\,km\,s$^{-1}$ and is red-shifted. We show all channel maps in figure~\ref{fig:channel maps}. The blue-shifted frequency channels clearly trace the northern tail detected in the SPHERE data, while the red shifted channels trace the tails to the West, in particular tail 1. Since we inferred from the scattered light data that tail 1 is above, and the northern tail below, the disk, we can thus conclude that both of these tails trace material falling onto the disk from the surrounding cloud. In order to check if the measured velocities are physical, we computed the free fall velocity around SU\,Aur and find values of 4-1.8\,km\,s$^{-1}$ for separation between 200\,au and 1200\,au. This is compatible with the projected velocities measured in the dust tails.\\
Additionally, we see strong red-shifted signal to the East and South-East of the Keplerian disk. If the detected scattered light signal is above the disk, then this may indicate that we see additional in-fall of material from these directions. It may also be that we simply trace the material falling in from the West as it is caught in Keplerian motion and spirals onto the disk.

\subsection{Possible foreground contamination}

In the red-shifted velocity channels there is in general emission with a velocity gradient from West to East, which may indicate that some of the signal is coming from the embedding cloud and not the infalling material. However, the blue-shifted channels are highly concentrated on the circumstellar disk and the northern tail, thus this is less of a concern in this case. We investigated if possible foreground contamination might be responsible for some of the signatures seen in scattered light. \\  
For the shadow lane feature to be produced by optically thick foreground material, we would require a thin filament aligned such that is crosses our line of sight towards the stellar position. To create a narrow feature like the shadow lane this filament would likely have to be close to the star. In this case we would expect to pick up an illumination signature of such a hypothetical structure in scattered light, in particular since we would see it under small scattering angles and thus close to the peak of the total intensity scattering phase function. Such a signature is not visible in the SPHERE or NACO data (neither in polarized nor total intensity). Additionally if such a structure was present, we would expected SU\,Aur to show significant extinction. The literature value for the the extinction of SU\,Aur is $A_V\sim$1\,mag \citep{Calvet2004} 
thus it does not appear that we observe the star obscured behind an optically thick filamentary structure. We therefore discard the possibility of highly localized foreground structures in front of the Keplerian disk and favor a misaligned inner disk scenario, supported by the orientation of the inner and outer disk derived by \cite{Labdon2019} and in this work. \\
Given the ALMA emission in the red-shifted channels, it may still be possible that thin, patchy foreground material influences the brightness distribution between the extended tail structures in scattered light.
In the red-shifted channels there is some emission visible at the position of the northern tail between velocities of 6.5 and 8.5 km/s. 
However the emission at the position of the norther tail is in all channels weaker than the emission located at tail 1 and tails 2-4. 
If a large portion of this red-shifted emission is foreground material, we thus conclude that the northern tail should be less affected than the region of tail 1-4. Nevertheless the northern tail is fainter in scattered light than tail 1, which we interpret as resulting from tail 1 being seen under small scattering angles and the northern tail under large scattering angles (i.e. tail 1 in front and the northern tail in the back). 
If significant foreground absorption is present, than the systematic should be such that tail 1 is even brighter compared to the northern tail if the absorption were not present. Thus it would not change our conclusion of the relative radial positions of the two structures as presented in section~\ref{scattering-tail-section}.

\subsection{Relative line fluxes in disk and tails}

\begin{figure}
\center
\includegraphics[width=0.50\textwidth]{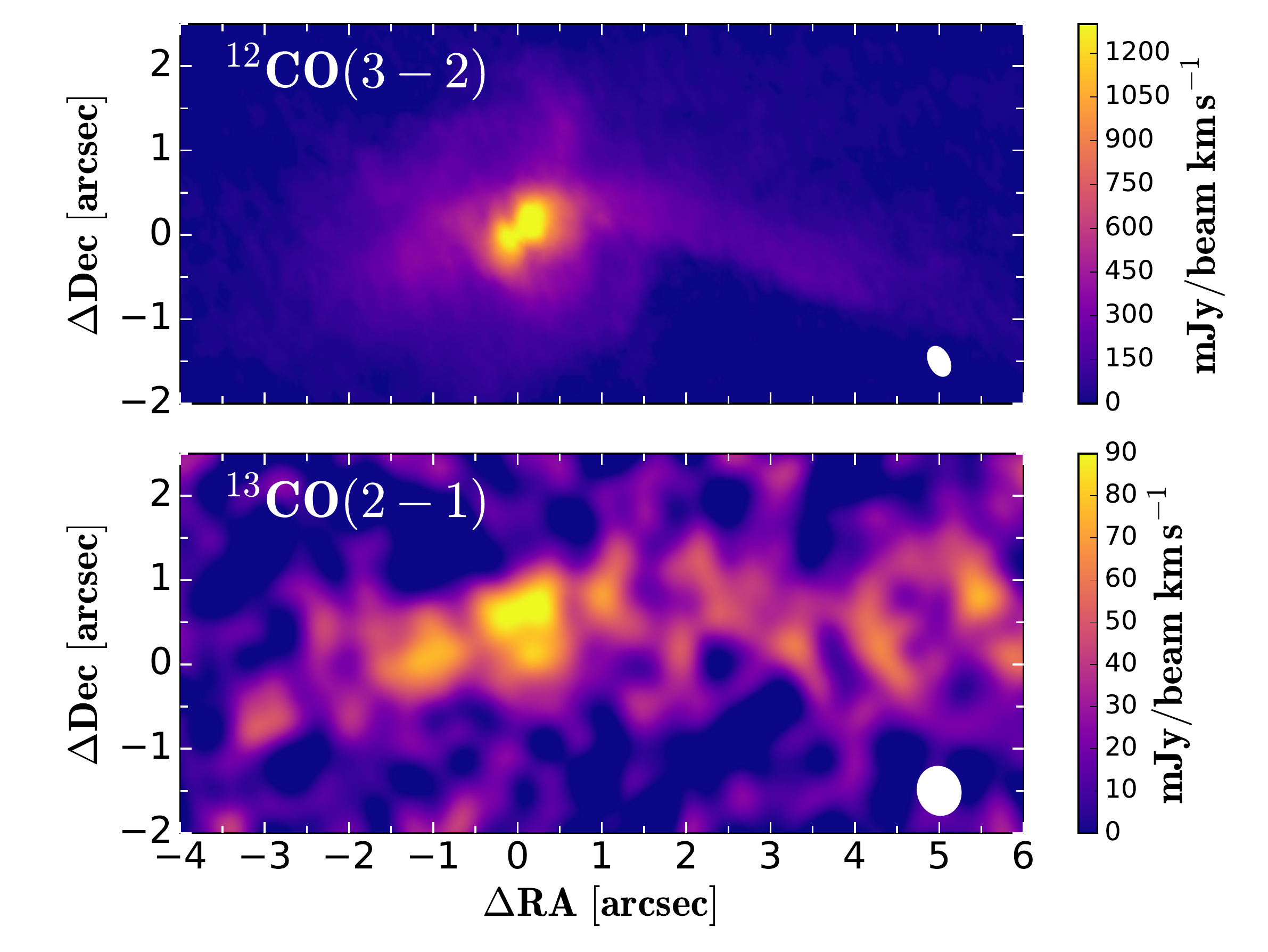}
\caption{Moment 0 map of the $^{12}$CO(3-2) and $^{13}$CO(2-1) data of the SU\,Aur system. The beam size is indicated by the white ellipse in the lower right corner.
} 
\label{fig: moment-maps}
\end{figure}

To test if the angular momentum transported by the dust tails is in principle sufficient to cause the misalignment of inner and outer disk that we discuss in section~\ref{sec: scatterd-light disk}, we used the available $^{12}$CO and $^{13}$CO line data to qualitatively assess the mass ratio between the Keplerian disk and the dust tails. Of the two line observations one expects $^{13}$CO to be optically thinner than $^{12}$CO and thus to trace more closely the density of the gas. \\
To estimate the mass ratio we integrated over the detected flux density in the moment 0 map of both data sets shown in figure~\ref{fig: moment-maps}.
For the Keplerian disk area we used a circular aperture with an outer radius of 0.7\arcsec{}. For the integrated flux in the tails we used two elliptical apertures in the $^{12}$CO data for the areas that coincide with the northern and the western tails in the SPHERE scattered light data. In the $^{13}$CO data we used only one elliptical aperture centered on the western region since the northern tail is not well detected in the $^{13}$CO emission. 
Using this procedure we find a flux density ratio of 0.6 for the $^{12}$CO data and 2.9 for the $^{13}$CO data.\\
Given that the Keplerian disk has a higher temperature than the tails, which are farther away from the star, a smaller column density of material is needed in the disk to produce the same amount of flux compared to the tails. For the optically thinner $^{13}$CO data we find roughly three times higher integrated flux in the tails than in the disk, which may indicate that there is significantly more mass in the tails than in the disk. We point out that this assumes a constant gas-to-dust ratio in both areas. Also even $^{13}$CO may still be optically thick in the disk (and possibly the tails) and thus we can not directly translate the flux density ratio to a mass ratio. 
However, it is encouraging that the flux density ratio between tail and disk increases between the optically thicker tracer $^{12}$CO and the thinner tracer $^{13}$CO. Given that the $^{13}$CO emission should be more density dominated than the $^{12}$CO emission this indicates that there is indeed a substantial amount of material in the tails compared to the disk.\\
For a proper measurement of the gas masses (and the angular momentum) in the different structures surrounding SU\,Aur deep observations of optically thin tracers are needed. However the measurements extracted from the existing data are well compatible with a scenario in which the infalling material misaligns the outer disk of SU\,Aur. 

\section{HST/STIS observations}

\begin{figure}
\center
\includegraphics[width=0.48\textwidth]{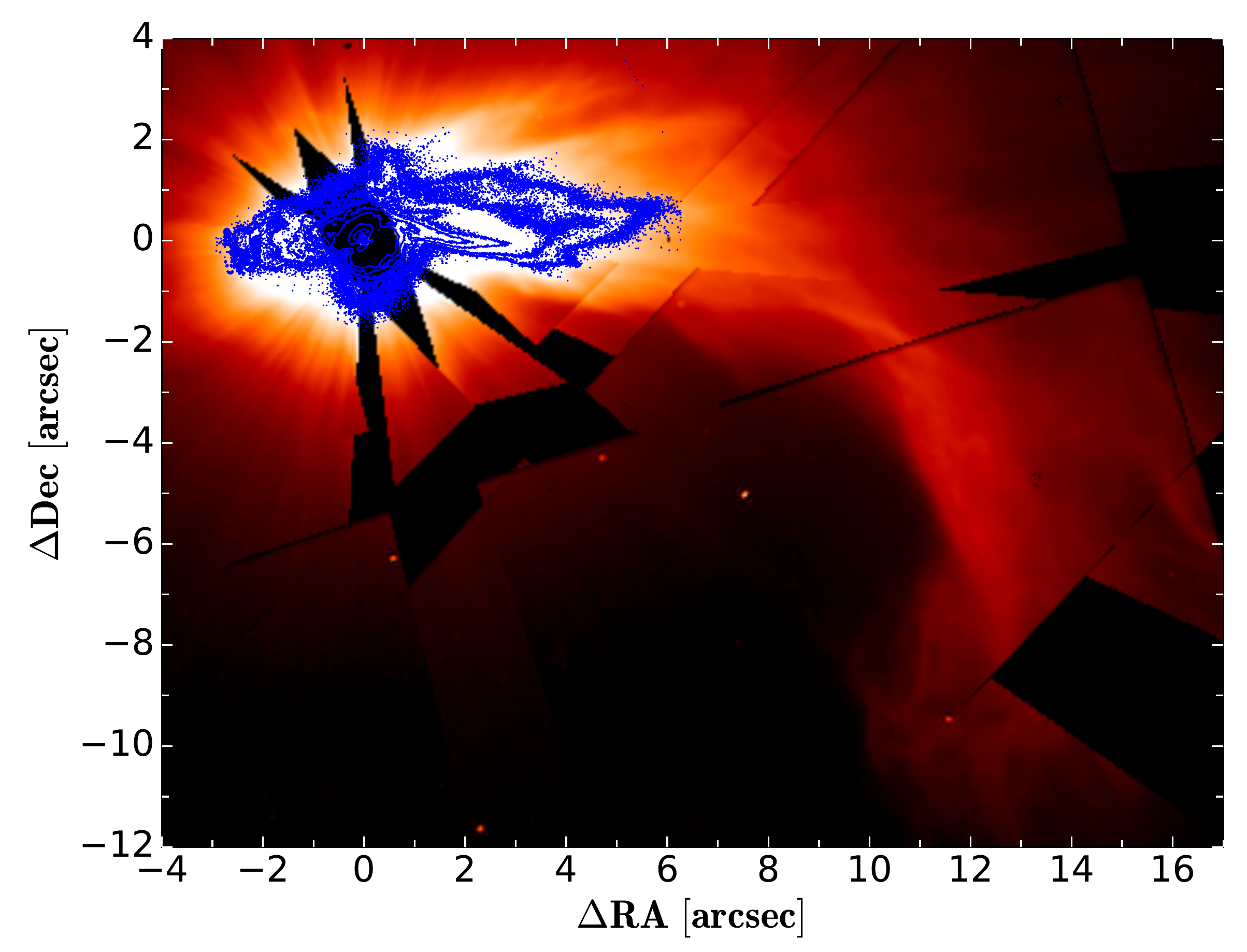} 
\caption{HST/STIS data first presented by \cite{2001AAS...199.6015G}. We masked the coronagraph and telescope spider features and combined two roll angles. SPHERE Q$\phi$ contours are overlayed in blue.
} 
\label{fig:stis-sphere}
\end{figure}

The large scale dust tail extending from SU Aur was previously captured in optical scattered light by \cite{2001AAS...199.6015G}, using HST/STIS. While in the STIS coronagraphic images the disk region is masked, STIS presents a significantly larger field of view. We show the HST/STIS data, after masking of the coronagraphic bars and combining of two telescope roll angles in figure~\ref{fig:stis-sphere}. We overlaid the contours of the SPHERE data for comparison, showing the complementarity of both instruments. \\
The dust tails seen in SPHERE seamlessly connect with the structures visible in the STIS image. In the STIS data it becomes apparent that the tail structure is turning towards the South. This is the direction in which AB\,Aur is located, which shows spectacular extended spiral features in scattered light \citep{Boccaletti2020}. The direction towards AB\,Aur may support the scenario discussed in \cite{Akiyama2019}, where they speculate that cloud material initially falls towards the center of gravity between SU\,Aur and AB\,Aur before it accretes onto either systems.  

\section{Discussion and conclusions}

The observational data of SU\,Aur paint an intricate picture of the formation of the system, its connection to the surrounding molecular cloud and its evolution.

\subsection{The origin of the dust tails}

SPHERE and ALMA data in concert show that material is falling onto the disk surrounding SU\,Aur. We can thus likely rule out a recent close encounter or the ejection of a dust clump for their origin \citep{Vorobyov2020}. 
Additionally, we checked the Gaia DR2 catalog and find no obvious candidates for a close stellar encounter with SU\,Aur within 100\arcsec{} (i.e. sources with similar distance or proper motion)\footnote{There are several known members of the Taurus\,X subgroup of which SU\,Aur is a member (\citealt{Luhman2009}), however the closest member is JH\,433, which is moving tangentially to SU\,Aur and thus is not a candidate for a close encounter. See appendix~\ref{Aur-cluster} for a brief overview.}, neither do we detect a point source within the 6\arcsec{} field of view of SPHERE (we are generally sensitive to all stellar and brown dwarf sources outside of $\sim$0.1\arcsec{}).\\  
The asymmetry of the tail structures from the largest scales seen with HST/STIS down to the smallest scales may suggest an interaction between SU\,Aur and the nearby young star AB\,Aur (projected separation of $\sim$29,000\,au). In \cite{Hacar2011} it is shown, that both of these systems are located in or near the same filament structure within the L\,1517 dark cloud.
This indicates that we may see the late formation of a very wide binary (or higher order) system formed by turbulent fragmentation (\citealt{Padoan1995}). The arc-like structure seen around AB\,Aur (\citealt{Grady1999}) and the large-scale dust and gas tails around SU\,Aur might be part of a connecting structure between the two systems as predicted by magneto-hydrodynamic simulations (e.g., \citealt{Kuffmeier2019}). As already suggested by \cite{Akiyama2019}, material may then be funneled along these structures towards either system.\\
The sharp tail structures in particular might be expected from classical Bondi-Hoyle accretion \citep{Bondi1944, Bondi1952}. 
Following \cite{Dullemond2019}, a cloudlet which undergoes a close encounter will form a large scale arc-like structure and possibly also sharper dust tails. This depends largely on the size of the cloudlet relative to the impact parameter of the encounter, but also the thermodynamics within the cloud. For a cloudlet with a radius larger than the impact parameter, which also cools efficiently, they produce scattered light images containing both large-scale arcs and smaller scale tails. Their synthetic scattered light images are reminiscent of the structure seen around SU\,Aur with the HST and the tails seen with SPHERE. Additionally, an encounter with a large cloudlet would also explain that we see not only red-shifted emission and scattered light in one direction but that it envelops the disk. \cite{Kuffmeier2020} also show that such close encounters with cloudlets can produce extended arcs on scales of 10$^4$\,au. Their simulations suggest that these resulting structures are long lived if the protostar is at rest relative to the surrounding gas and is encountered by a cloudlet in relative motion. This may be plausible for SU\,Aur for two reasons. On the one hand, the fact that we indeed detect these dynamical signatures is itself an indication that they are long lived. On the other hand the systemic velocity of $\sim$6\,kms$^{-1}$ fits well with the radial velocity of the surrounding filament as reported in \cite{Hacar2011}.\\

\subsection{Disk instability due to infalling material?}

The new SPHERE observations allow us to trace the large scale structures in SU\,Aur seamlessly down to scales of less than 10\,au. The two most striking features in the Keplerian disk are the multitude of spiral arms and the sharp shadow lane. Both of these can well be explained by the infall of material. Spiral waves are a common consequence of instability triggered by infalling material (\citealt{Moeckel2009, Lesur2015, Bae2015, Kuffmeier2018, Dullemond2019, Kuffmeier2020}). While simulations typically show the spiral features in the gas, we can expect to trace them in scattered light, since small dust particles are well coupled to the gas. Indeed most of these simulations produce disks with a large number of "wispy" spiral features that match closely the appearance of SU\,Aur as highlighted in figure~\ref{fig:sphere-tails}. Such spiral features have so far been observed in scattered light predominantly in circumbinary disks, i.e., HD\,142527 (\citealt{Avenhaus2017}), HD\,34700 (\citealt{Monnier2019}) and GG\,Tau (\citealt{Keppler2020}). At this time there is no evidence that SU\,Aur is a binary star. In particular it lacks the large central cavity seen in the other systems. The spiral structure rather resembles the one seen in AB\,Aur by \cite{Boccaletti2020}. This seems to fit in the picture of asymmetric late infall in both systems along the embedding filament. However, we note that \cite{Poblete2020} point out that the inner spiral structure in the disk around AB\,Aur could be caused by a stellar binary.\\

\subsection{Disk misalignment by late infall?}

The shadow lane in SU\,Aur is a feature now commonly seen in scattered light images (e.g., \citealt{Marino2015,Stolker2016a,Benisty2017,Keppler2020}) and typically explained by a misalignment or warp between inner and outer disk. Indeed by comparing our ALMA continuum fit with the interferometric result from \cite{Labdon2019} we find a relative misalignment of $\sim$70$^\circ$.
While there is ample theory on how such a misalignment is caused, there is little observational evidence of the process. \cite{Brinch2016} reported on the misalignment of circumstellar disks in the IRS\,43 multiple system with respect to the surrounding circumbinary disk, presumably caused by the chaotic interaction of the stellar cores. An even more spectacular case of such a misalignment by multiple stars was recently shown for for the GW\,Ori system by \cite{Bi2020} and \cite{Kraus2020}. 
\cite{Sakai2019} found a warped disk around the proto-star IRAS\,04368+2557. They inferred from the absence of signs of a close stellar encounter that the warp should be caused by late infall of material, but did not find direct evidence. For AB\,Aur, \cite{Tang2012} show a large scale warp in the surrounding disk and multiple tentative spiral features and suggest that both are caused by late infall. Given the results by \cite{Boccaletti2020}, who detect large scale spiral structures in scattered light, down to disk scales, this seems a likely scenario. However, we note that \cite{Boccaletti2020} interpret the innermost spiral structures as signs of a forming proto-planet rather than as instability caused by infalling material.
In SU\,Aur we directly detect the infalling material and can trace it from thousands of au down to disk scales. As \cite{Dullemond2019} argue, infalling material is bound to have a vastly different orientation of angular momentum compared to the accreting disk. In section~\ref{section: ALMA comparion}, we discussed that the mass estimates in the tail and disk structure make such a scenario plausible. Thus the infall we trace is likely causing a warp of the outer disk regions. This makes the AB\,Aur and SU\,Aur pair the best examples of such effects caused by late infall. We note that it is in principle possible that the disk we currently see in scattered light around SU\,Aur is not primordial at all, but is actually formed as a result of a close encounter with a cloudlet (see \citealt{Dullemond2019}). In this case it would be natural that it is misaligned with respect to the (presumably) primordial inner disk detected by \cite{Labdon2019}.\\ 
The structures revealed around SU\,Aur by SPHERE and ALMA form a coherent picture of late infall of material that dominates the evolution of the protoplanetary disk. This mechanism not only provides an additional mass reservoir for forming planets (see the discussion in \citealt{Manara2018}) but can also trigger planet formation by gravitational instability. As suggested by \cite{Thies2011}, this scenario might be able to explain the spin-orbit misalignment found in evolved planetary systems. These new high-resolution observations enable detailed future simulations of such planet formation pathways.  

\acknowledgments

We thank an anonymous referee for a thorough review that improved the paper.
We would like to thank Jonathan Williams and Antonio Garufi for fruitful discussion. We also thank Eiji Akiyama for providing their reduced ALMA data and Aaron Labdon and the A\&A journal for authorizing the reprint of the near infrared interferometric results. 

SPHERE is an instrument designed and built by a consortium
consisting of IPAG (Grenoble, France), MPIA (Heidelberg, Germany), LAM (Marseille, France), LESIA (Paris, France), Laboratoire Lagrange (Nice, France), INAF - Osservatorio di Padova (Italy), Observatoire de
Gen\`{e}ve (Switzerland), ETH Zurich (Switzerland), NOVA (Netherlands), ONERA
(France), and ASTRON (The Netherlands) in collaboration with ESO.
SPHERE was funded by ESO, with additional contributions from CNRS
(France), MPIA (Germany), INAF (Italy), FINES (Switzerland), and NOVA
(The Netherlands). SPHERE also received funding from the European Commission
Sixth and Seventh Framework Programmes as part of the Optical Infrared
Coordination Network for Astronomy (OPTICON) under grant number RII3-Ct2004-001566
for FP6 (2004-2008), grant number 226604 for FP7 (2009-2012),
and grant number 312430 for FP7 (2013-2016).

C.G. acknowledges funding from the Netherlands Organisation for Scientific Research (NWO) TOP-1 grant as part
of the research program “Herbig Ae/Be stars, Rosetta stones for understanding
the formation of planetary systems”, project number 614.001.552.
Support for this work was provided by NASA through the NASA Hubble Fellowship grant \#HST-HF2-51460.001-A awarded by the Space Telescope Science Institute, which is operated by the Association of Universities for Research in Astronomy, Inc., for NASA, under contract
NAS5-26555.
This paper makes use of ALMA data ADS/JAO.ALMA\#2013.1.00426.S. ALMA is a partnership of ESO (representing its member states), NSF (USA) and NINS (Japan), together with NRC (Canada) and NSC and ASIAA (Taiwan), in cooperation with the Republic of Chile. The Joint ALMA Observatory is operated by ESO, AUI/NRAO and NAOJ. The National Radio Astronomy Observatory is a facility of the National Science Foundation operated under cooperative agreement by Associated Universities, Inc. We thank the North American ALMA Science Center for providing resources to reduce ALMA data.
T.B. acknowledges funding from the European Research Council under the European Union’s Horizon 2020 research and innovation programme under grant agreement No 714769 and funding from the Deutsche Forschungsgemeinschaft under Ref. no. FOR 2634/1 and under Germany's Excellence Strategy (EXC-2094–390783311).
JB acknowledges support by NASA through the NASA Hubble Fellowship grant \#HST-HF2-51427.001-A awarded  by  the  Space  Telescope  Science  Institute,  which  is  operated  by  the  Association  of  Universities  for  Research  in  Astronomy, Incorporated, under NASA contract NAS5-26555.
This research has used the SIMBAD database, operated at CDS, Strasbourg, France \citep{Wenger2000}. 
Part of this research was carried out at the Jet Propulsion
Laboratory, California Institute of Technology, under a contract with
the National Aeronautics and Space Administration (80NM0018D0004). EEM acknowledges support from the Jet Propulsion
Laboratory Exoplanetary Science Initiative, NASA award 17-K2GO6-0030, and NASA grant NNX15AD53G.
We used the \emph{Python} programming language\footnote{Python Software Foundation, \url{https://www.python.org/}}, especially the \emph{SciPy} \citep{2020SciPy-NMeth}, \emph{NumPy} \citep{oliphant2006guide}, \emph{Matplotlib} \citep{Matplotlib} and \emph{astropy} \citep{astropy_1,astropy_2} packages.
We thank the writers of these software packages for making their work available to the astronomical community.

%

\vspace{2mm}
\facilities{VLT(SPHERE), VLT(NACO), ALMA, HST(STIS)}





\appendix

\section{Iterative feedback reference and angular differential imaging}
\label{IRDI-IADI}

\begin{figure*}[h!]
\center
\includegraphics[width=0.98\textwidth]{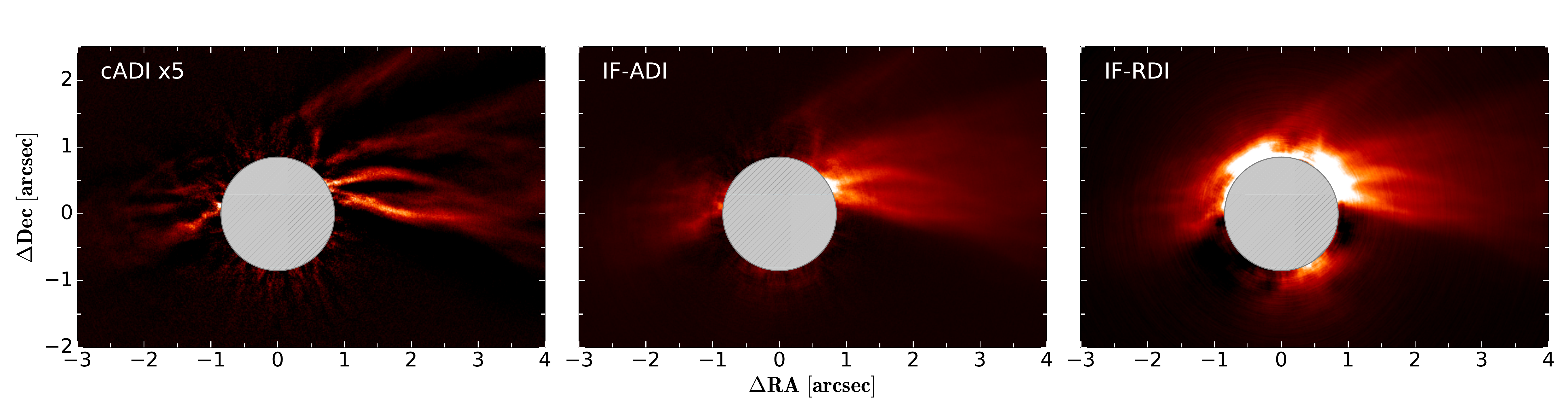} 
\caption{Total intensity images of SU\,Aur after post-processing of the SPHERE H-band data. \textit{Left:} Classical angular differential imaging reduction. \textit{Center:} Iterative feedback angular differential imaging reduction of the same data set. The result after 100 iterations is shown. \textit{Right:} Iterative feedback reference differential imaging reduction. The result after 14 iterations is shown. All panels are shown in the same linear color scale. The cADI reduction was multiplied by a factor of 5 to improve the contrast in this color scale. The innermost region is not well recovered in all cases an thus masked out.} 
\label{fig:intensity-compare}
\end{figure*}

The total intensity image was extracted from the data by using an iterative reference differential imaging (RDI) approach. In our RDI routine, scaled reference PSF images were created by projecting the data on a reference star library, using the KLIP algorithm as introduced by \cite{Soummer2012}. Furthermore, an iterative approach was applied to this KLIP RDI routine, with the aim to reduce overestimation of residual stellar light in the scaled reference PSF caused by disk signal in the data affecting the projection on the reference star library. This approach, named Iterative Disk Feedback (IDF), will be presented in a forthcoming publication by Vaendel et al. (in prep.), who show it to effectively minimize over-subtraction of stellar signal in the RDI reduction of data containing extended disk structures. The iterative approach enables to subtract disk signal (selected from the resulting reduction of the previous iteration) from the image stack which is used for projection on the reference star library in the regular KLIP RDI routine. This means that for each iteration, the KLIP RDI reduction is less affected by the presence of (identified) disk signal. Ideally, the iterative process continues until all disk signal is identified and hence only the actual stellar signal in the data projected onto the reference star library for the creation of the scaled reference PSF. In our case the reduction did not significantly improve anymore after 14 iterations, where we stopped the IDF process.\\
Iterative Angular Differential Imaging uses median subtraction ADI multiple times over on the
same dataset. It takes the final result coming out of the ADI pipeline and subtracts all positive signal
of this result from the original dataset after it has been rotated by the respective field rotation of
each of the images. By setting the negative values to zero the self-subtracted regions are not fed
back. In this way, less signal of the disk is present in the dataset. When computing the median of the
dataset, less of the disk signal is also present in the resulting PSF. Consequently, less self-
subtraction of the disk occurs when the median is subtracted. However, after computing the PSF,
there is still some disk signal left in the dataset after one iteration. The more iterations, the less
signal of the disk is left in the PSF and thus the better the final result is. For the dataset used in this
work, this iterative process was done 100 times over. With this technique some problems are
occurring. After many iterations the star is reconstructed and ring-like structures are being generated
due to the many rotations happening during the iteration process. These effects mainly occur at the
centre of the image due to most signal being present there. In our work, the main science occurs at
the outer dust tails, where these effects are not dominating over the disk signal.
We show the resulting total intensity images for all techniques in figure~\ref{fig:intensity-compare}.

\section{Gaussian fit of the ALMA visibilities}
\label{app:alma-fit}

In order to derive inclination and position angle of the mm continuum disk observed by ALMA we perform a fit of the visibilities in the $uv$ plane. The continuum visibilities of all spectral windows have been channel averaged to 250\,MHz wide channels to reduce the amount of data used in the fitting, without inducing any bandwidth smearing. We assumed a simple radial Gaussian profile for the intensity distribution, with intensity normalization $I_1$ and Gaussian width $\sigma_1$ as free parameters. Additional four parameters are accounted for in the fitting, in particular inclination, position angle, and disk center. The Fourier transform of the intensity profile is generated using the \texttt{GALARIO} code \citep{Tazzari2018}, with the same sampling as in the ALMA observations. The parameter space is then explored using the \texttt{emcee} package \citep{emcee}, assuming wide uniform priors on all parameters. The posterior distribution is sampled with 60 walkers for 4000 steps after a burn in of 1000 steps. 

\begin{figure}[h!]
\center
\includegraphics[width=0.6\textwidth]{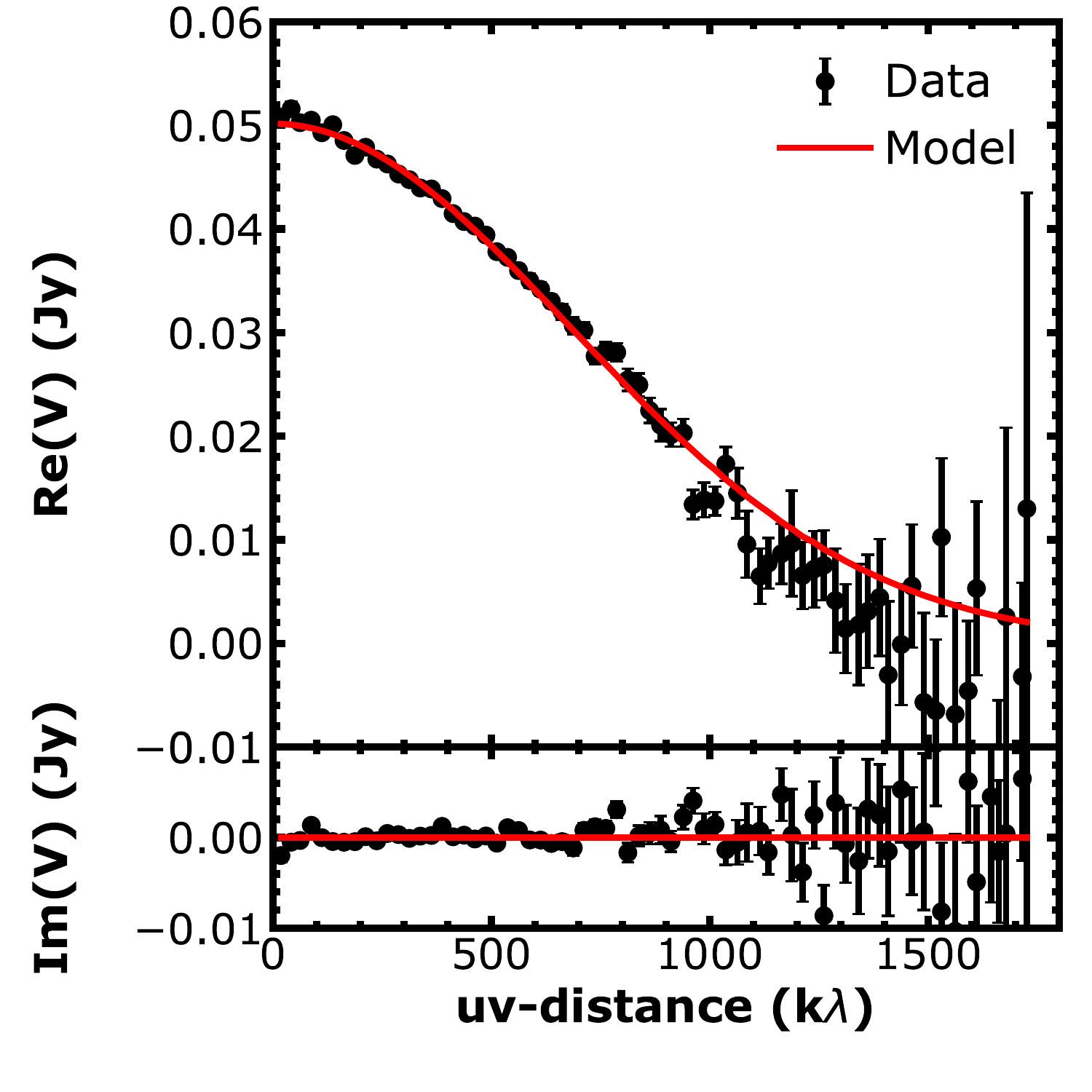} 
\caption{Visibilities of the ALMA continuum data, re-centered and de-projected using the best fit geometrical parameters described in section~\ref{app:alma-fit}. The disk is well resolved and well described by a simple Gaussian approximation.} 
\label{fig:alma-visibilities}
\end{figure}

\begin{figure}[h!]
\center
\includegraphics[width=0.98\textwidth]{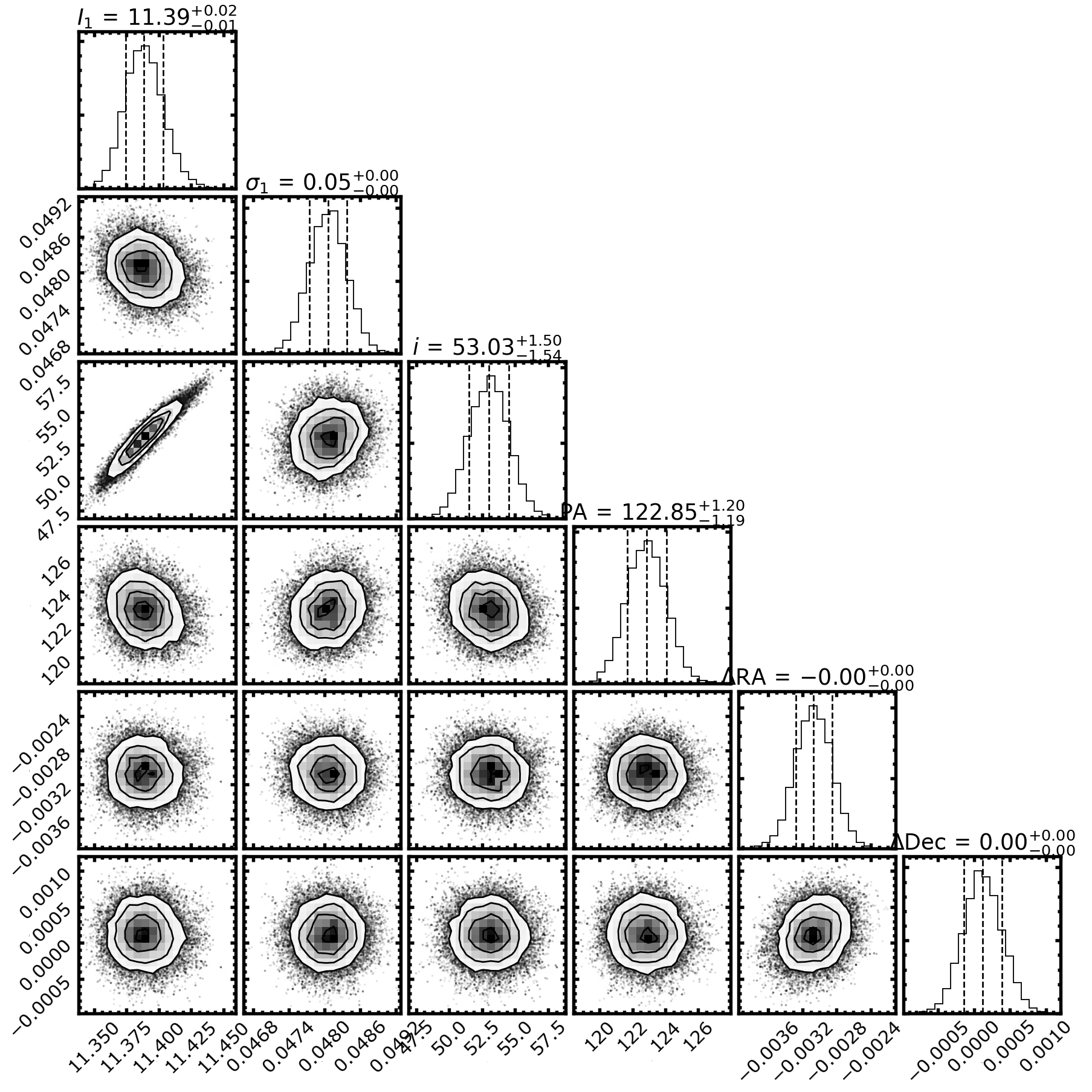} 
\caption{Marginalized posterior distributions of MCMC fitting the continuum visibilities of the ALMA data.} 
\label{fig:alma-visibilities-results}
\end{figure}

The fit nicely converges to a solution, where the disk is clearly well resolved and well reproduced by the simple Gaussian model (see figure~\ref{fig:alma-visibilities}). Table \ref{tab:best_fit_alma} reports the best fit parameters, taken as the median value of the marginalized posterior distributions shown in figure~\ref{fig:alma-visibilities-results}. No clear structure is seen in the residuals above the $3\sigma$ limit. Alternative parametrization for the intensity profile (as broken power laws or modified versions of it) did not improve the results.

\begin{table*}[h!]
\centering
\begin{tabular}{cccccc}
\hline
$I_1$ & $\sigma_1$	& $i$	& PA	& $\Delta$RA 	& $\Delta$Dec	\\
($\log\,$Jy/steradian) & ($\arcsec$) & $(^{\circ})$ & $(^{\circ})$ & (mas) & (mas) \\
\hline
$11.39^{+0.02}_{-0.01}$ & $0.048^{+0.001}_{-0.001}$ & $53.0^{+1.5}_{-1.5}$ & $122.8^{+1.2}_{-1.2}$ & $-3.1^{+0.2}_{-0.2}$ & $0.1^{+0.3}_{-0.3}$ \\
\hline
\hline
\end{tabular}
\label{tab:best_fit_alma}
\caption{Median of the marginalized posteriors of the fitted parameters for the continuum emission of SU Aur, with associated statistical uncertainties from  the 16th and 84th percentiles of the marginalized distributions.}
\end{table*}

\section{SPHERE data and ALMA overlays}
\label{ALMA-appendix}

\begin{figure}[h!]
\center
\includegraphics[width=0.98\textwidth]{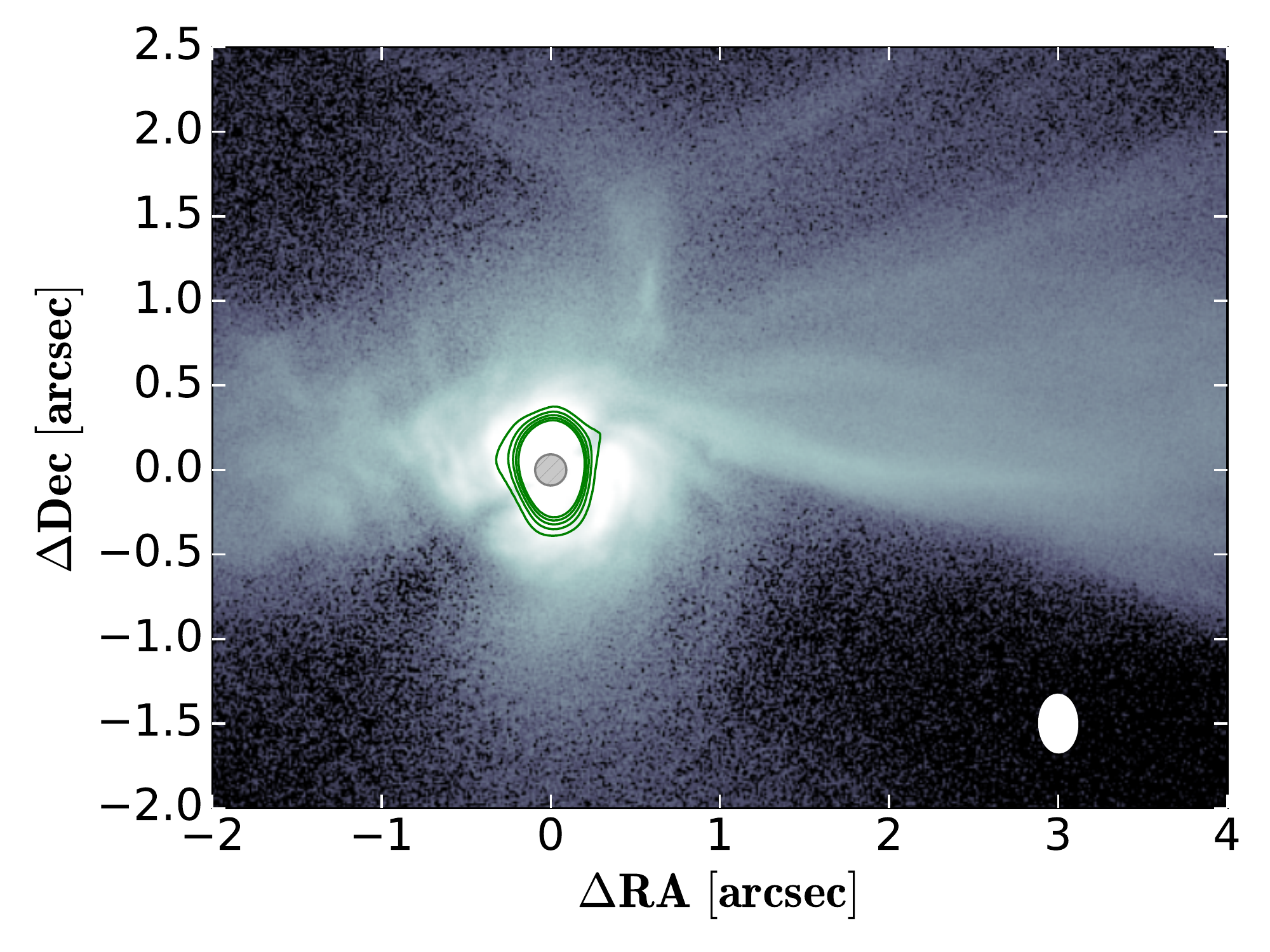} 
\caption{SPHERE Q$_\phi$ image with the ALMA dust continuum emission in Band 7 overlayed in green contours. The ALMA beam size and orientation is indicated by the white ellipse in the lower right corner. Contour levels are 4,8,12,16 and 20\,$\sigma$.
} 
\label{fig:sphere-alma-dust}
\end{figure}

\begin{figure*}[h!]
\center
\includegraphics[width=0.98\textwidth]{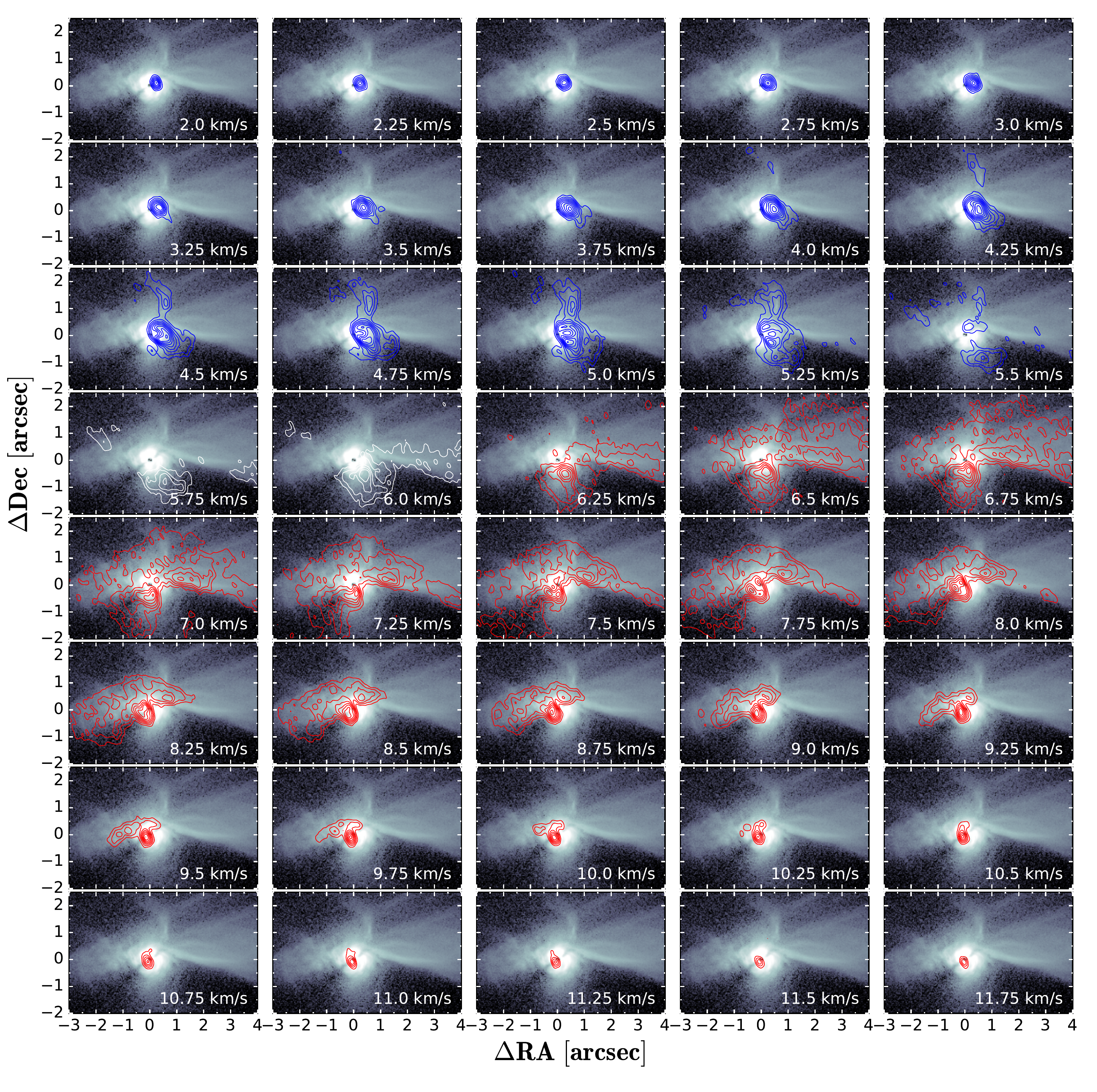} 
\caption{SPHERE Q$_\phi$ image with the ALMA $^{12}$CO 3-2 channel maps as contours. We show contours of channels blue-shifted relative to the systemic velocity of $\sim$6\,kms$^{-1}$ in blue, red-shifted channels in red and the channels closest to the systemic velocity in white.
The lowest contour represents 5\,$\sigma$ and subsequent contours are steps of 2\,$\sigma$.} 
\label{fig:channel maps}
\end{figure*}


\section{Cluster members near SU\,Aur}
\label{Aur-cluster}

\begin{figure}[h!]
\center
\includegraphics[width=0.68\textwidth]{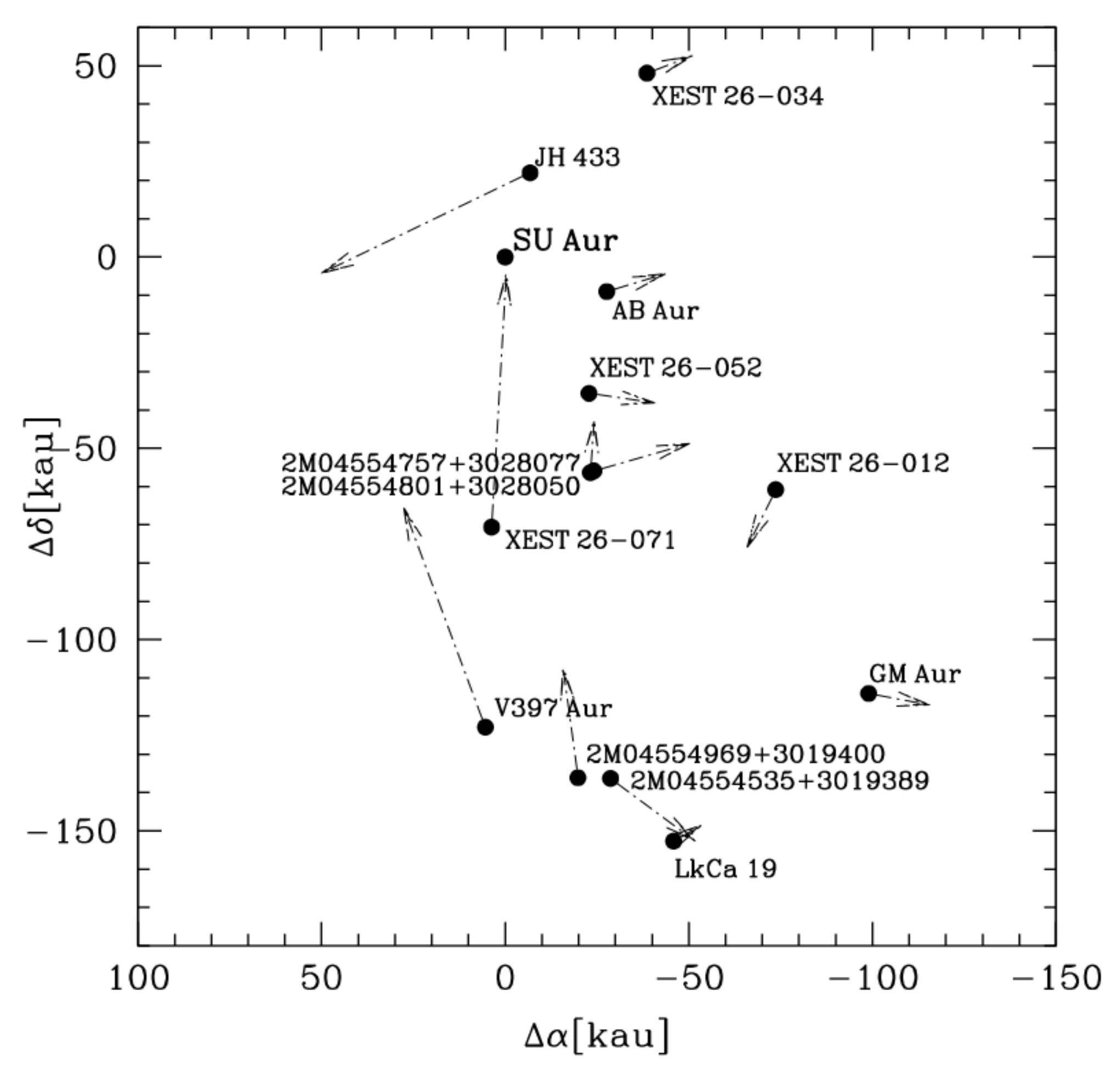} 
\caption{Known group members of the Taurus\,X group from \cite{Luhman2009}. The proper motion in the plane of the sky is indicated by the arrows and points to the projected position 140\,kyr in the future.} 
\label{fig:taurusx}
\end{figure}

SU Aur is located in a small embedded cluster associated with the L1517 cloud (Taurus\,X in \citealt{Luhman2009}). The distribution of members of this subgroup is studied as one of the "NESTs" (\#20) in \cite{Joncour2018}. They list 13 members in Gaia DR2. The other cluster members are all within a projected separation of 0.77\,pc of SU\,Aur. The closest members to SU Aur are JH 433 (146\arcsec{}, 23\,kau), AB\,Aur (184\arcsec{}, 29\,kau), and XEST 26-052 (268\arcsec{}, 42\,kAU). The mean separation between the members of the group is $\sim$25\,kAU (\citealt{Joncour2018}). In figure~\ref{fig:taurusx} we show the 2d distribution of group members and indicate their proper motions.

\bibliography{MyBibFMe}{}
\bibliographystyle{aasjournal}

\end{document}